\documentclass[journal]{IEEEtran} %
\usepackage{subfig}
\usepackage[utf8]{inputenc} 
\usepackage{graphicx,url}
\usepackage{booktabs}       
\graphicspath{ {./images/} }
\usepackage{multirow}
\usepackage{hhline}
\usepackage{graphicx}
\usepackage{amssymb,amsmath,mathtools}
\usepackage{amsthm}

\usepackage{bm}                 %

\usepackage{color,soul}
\usepackage[dvipsnames]{xcolor}
\usepackage{bbm}
\usepackage{lipsum,lineno}
\usepackage{float}
\makeatletter
\newif\if@restonecol
\makeatother
\usepackage{amsmath}
\usepackage{kbordermatrix}
\usepackage{mathrsfs}
\usepackage[shortlabels]{enumitem}
\usepackage{algorithm}
\usepackage{algpseudocode}
\usepackage{fancyhdr} 
\usepackage{tabularx}
\usepackage{dsfont}
\usepackage{comment} 

\usepackage{tikz,pgfplots}
\usepackage{pgfplots}
\usetikzlibrary{shapes.geometric,backgrounds,patterns, trees}
\usetikzlibrary{3d,decorations.text,shapes.arrows,positioning,fit,backgrounds}
\usetikzlibrary{positioning, decorations.pathmorphing, shapes}
\usetikzlibrary{decorations.pathreplacing}
\usetikzlibrary{shapes.geometric,backgrounds,patterns, trees}
\usetikzlibrary{arrows.meta,
                bending,
                intersections,
                quotes,
                shapes.geometric}
              \usetikzlibrary{automata, positioning}
              \usepgfplotslibrary{fillbetween}
\pgfmathdeclarefunction{gauss}{2}{%
  \pgfmathparse{1/(#2*sqrt(2*pi))*exp(-((x-#1)^2)/(2*#2^2))}%
}
\usetikzlibrary{shapes,arrows}
\usetikzlibrary{arrows.meta}
\usetikzlibrary{positioning}
\tikzset{set/.style={draw,circle,inner sep=0pt,align=center}}
\usetikzlibrary{automata, positioning}
  \usetikzlibrary{shapes,shadows}
  \tikzstyle{abstractbox} = [draw=black, fill=white, rectangle,
  inner sep=10pt, style=rounded corners, drop shadow={fill=black,
  opacity=1}]
\tikzstyle{abstracttitle} =[fill=white]
\usetikzlibrary{calc,positioning,shapes.geometric}
\usetikzlibrary{arrows.meta,arrows}

\usetikzlibrary{matrix}
\tikzstyle{cblue}=[circle, draw, thin,fill=cyan!20, scale=0.8]
\tikzstyle{qgre}=[rectangle, draw, thin,fill=green!20, scale=0.8]
\tikzstyle{rpath}=[ultra thick, red, opacity=0.4]
\tikzstyle{legend_isps}=[rectangle, rounded corners, thin,
                       fill=gray!20, text=blue, draw]

\tikzstyle{legend_overlay}=[rectangle, rounded corners, thin,
                           top color= white,bottom color=green!25,
                           minimum width=2.5cm, minimum height=0.8cm,
                           pinegreen]
\tikzstyle{legend_phytop}=[rectangle, rounded corners, thin,
                          top color= white,bottom color=cyan!25,
                          minimum width=2.5cm, minimum height=0.8cm,
                          royalblue]
\tikzstyle{legend_general}=[rectangle, rounded corners, thin,
                          top color= white,bottom color=lavander!25,
                          minimum width=2.5cm, minimum height=0.8cm,
                          violet]
                          \usetikzlibrary{matrix}
\def\layersep{2cm}
\colorlet{myRed}{red!20}
\tikzset{
  rows/.style 2 args={/utils/temp/.style={row ##1/.append style={nodes={#2}}},
    /utils/temp/.list={#1}},
  columns/.style 2 args={/utils/temp/.style={column ##1/.append style={nodes={#2}}},
    /utils/temp/.list={#1}}}
\usetikzlibrary{backgrounds,calc,shadings,shapes.arrows,shapes.symbols,shadows}
\definecolor{switch}{HTML}{006996}

\makeatletter
\pgfkeys{/pgf/.cd,
  parallelepiped offset x/.initial=2mm,
  parallelepiped offset y/.initial=2mm
}
\pgfdeclareshape{parallelepiped}
{
  \inheritsavedanchors[from=rectangle] 
  \inheritanchorborder[from=rectangle]
  \inheritanchor[from=rectangle]{north}
  \inheritanchor[from=rectangle]{north west}
  \inheritanchor[from=rectangle]{north east}
  \inheritanchor[from=rectangle]{center}
  \inheritanchor[from=rectangle]{west}
  \inheritanchor[from=rectangle]{east}
  \inheritanchor[from=rectangle]{mid}
  \inheritanchor[from=rectangle]{mid west}
  \inheritanchor[from=rectangle]{mid east}
  \inheritanchor[from=rectangle]{base}
  \inheritanchor[from=rectangle]{base west}
  \inheritanchor[from=rectangle]{base east}
  \inheritanchor[from=rectangle]{south}
  \inheritanchor[from=rectangle]{south west}
  \inheritanchor[from=rectangle]{south east}
  \backgroundpath{
    \southwest \pgf@xa=\pgf@x \pgf@ya=\pgf@y
    \northeast \pgf@xb=\pgf@x \pgf@yb=\pgf@y
    \pgfmathsetlength\pgfutil@tempdima{\pgfkeysvalueof{/pgf/parallelepiped
      offset x}}
    \pgfmathsetlength\pgfutil@tempdimb{\pgfkeysvalueof{/pgf/parallelepiped
      offset y}}
    \def\ppd@offset{\pgfpoint{\pgfutil@tempdima}{\pgfutil@tempdimb}}
    \pgfpathmoveto{\pgfqpoint{\pgf@xa}{\pgf@ya}}
    \pgfpathlineto{\pgfqpoint{\pgf@xb}{\pgf@ya}}
    \pgfpathlineto{\pgfqpoint{\pgf@xb}{\pgf@yb}}
    \pgfpathlineto{\pgfqpoint{\pgf@xa}{\pgf@yb}}
    \pgfpathclose
    \pgfpathmoveto{\pgfqpoint{\pgf@xb}{\pgf@ya}}
    \pgfpathlineto{\pgfpointadd{\pgfpoint{\pgf@xb}{\pgf@ya}}{\ppd@offset}}
    \pgfpathlineto{\pgfpointadd{\pgfpoint{\pgf@xb}{\pgf@yb}}{\ppd@offset}}
    \pgfpathlineto{\pgfpointadd{\pgfpoint{\pgf@xa}{\pgf@yb}}{\ppd@offset}}
    \pgfpathlineto{\pgfqpoint{\pgf@xa}{\pgf@yb}}
    \pgfpathmoveto{\pgfqpoint{\pgf@xb}{\pgf@yb}}
    \pgfpathlineto{\pgfpointadd{\pgfpoint{\pgf@xb}{\pgf@yb}}{\ppd@offset}}
  }
}

\makeatletter
\tikzset{anchor/.append code=\let\tikz@auto@anchor\relax,
  add font/.code=%
    \expandafter\def\expandafter\tikz@textfont\expandafter{\tikz@textfont#1},
  left delimiter/.style 2 args={append after command={\tikz@delimiter{south east}
    {south west}{every delimiter,every left delimiter,#2}{south}{north}{#1}{.}{\pgf@y}}}}
\tikzstyle{sms} = [rectangle callout, draw,very thick, rounded corners, minimum height=20pt]
\makeatletter
\tikzset{anchor/.append code=\let\tikz@auto@anchor\relax,
  add font/.code=%
    \expandafter\def\expandafter\tikz@textfont\expandafter{\tikz@textfont#1},
  left delimiter/.style 2 args={append after command={\tikz@delimiter{south east}
    {south west}{every delimiter,every left delimiter,#2}{south}{north}{#1}{.}{\pgf@y}}}}
\tikzstyle{sms} = [rectangle callout, draw,very thick, rounded corners, minimum height=20pt]
\usetikzlibrary{positioning,calc}

\tikzset{
  mybackground9/.style={execute at end picture={
        \begin{scope}[on background layer]
          \draw[black,fill=black!5,rounded corners=6ex] (current bounding box.south west)
                    rectangle (current bounding box.north east);
          \node[draw,fill=white,ellipse,anchor=west,inner sep=1pt,minimum width=4ex] at (current bounding box.north
                   west){#1};
        \end{scope}
    }},
}

\tikzset{l3 switch/.style={
    parallelepiped,fill=switch, draw=white,
    minimum width=0.75cm,
    minimum height=0.75cm,
    parallelepiped offset x=1.75mm,
    parallelepiped offset y=1.25mm,
    path picture={
      \node[fill=white,
        circle,
        minimum size=6pt,
        inner sep=0pt,
        append after command={
          \pgfextra{
            \foreach \angle in {0,45,...,360}
            \draw[-latex,fill=white] (\tikzlastnode.\angle)--++(\angle:2.25mm);
          }
        }
      ]
       at ([xshift=-0.75mm,yshift=-0.5mm]path picture bounding box.center){};
    }
  },
  ports/.style={
    line width=0.3pt,
    top color=gray!20,
    bottom color=gray!80
  },
  rack switch/.style={
    parallelepiped,fill=white, draw,
    minimum width=1.25cm,
    minimum height=0.25cm,
    parallelepiped offset x=2mm,
    parallelepiped offset y=1.25mm,
    xscale=-1,
    path picture={
      \draw[top color=gray!5,bottom color=gray!40]
      (path picture bounding box.south west) rectangle
      (path picture bounding box.north east);
      \coordinate (A-west) at ([xshift=-0.2cm]path picture bounding box.west);
      \coordinate (A-center) at ($(path picture bounding box.center)!0!(path
        picture bounding box.south)$);
      \foreach \x in {0.275,0.525,0.775}{
        \draw[ports]([yshift=-0.05cm]$(A-west)!\x!(A-center)$)
          rectangle +(0.1,0.05);
        \draw[ports]([yshift=-0.125cm]$(A-west)!\x!(A-center)$)
          rectangle +(0.1,0.05);
       }
      \coordinate (A-east) at (path picture bounding box.east);
      \foreach \x in {0.085,0.21,0.335,0.455,0.635,0.755,0.875,1}{
        \draw[ports]([yshift=-0.1125cm]$(A-east)!\x!(A-center)$)
          rectangle +(0.05,0.1);
      }
    }
  },
  server/.style={
    parallelepiped,
    fill=white, draw,
    minimum width=0.35cm,
    minimum height=0.75cm,
    parallelepiped offset x=3mm,
    parallelepiped offset y=2mm,
    xscale=-1,
    path picture={
      \draw[top color=gray!5,bottom color=gray!40]
      (path picture bounding box.south west) rectangle
      (path picture bounding box.north east);
      \coordinate (A-center) at ($(path picture bounding box.center)!0!(path
        picture bounding box.south)$);
      \coordinate (A-west) at ([xshift=-0.575cm]path picture bounding box.west);
      \draw[ports]([yshift=0.1cm]$(A-west)!0!(A-center)$)
        rectangle +(0.2,0.065);
      \draw[ports]([yshift=0.01cm]$(A-west)!0.085!(A-center)$)
        rectangle +(0.15,0.05);
      \fill[black]([yshift=-0.35cm]$(A-west)!-0.1!(A-center)$)
        rectangle +(0.235,0.0175);
      \fill[black]([yshift=-0.385cm]$(A-west)!-0.1!(A-center)$)
        rectangle +(0.235,0.0175);
      \fill[black]([yshift=-0.42cm]$(A-west)!-0.1!(A-center)$)
        rectangle +(0.235,0.0175);
    }
  },
}

\usetikzlibrary{calc, shadings, shadows, shapes.arrows}

\tikzset{%
  interface/.style={draw, rectangle, rounded corners, font=\LARGE\sffamily},
  ethernet/.style={interface, fill=yellow!50},
  serial/.style={interface, fill=green!70},
  speed/.style={sloped, anchor=south, font=\large\sffamily},
  route/.style={draw, shape=single arrow, single arrow head extend=4mm,
    minimum height=1.7cm, minimum width=3mm, white, fill=switch!20,
    drop shadow={opacity=.8, fill=switch}, font=\tiny}
}
\newcommand*{\shift}{1.3cm}
\newcommand{\Crossk}{$\mathbin{\tikz [x=1.2ex,y=1.2ex,line width=.1ex, black] \draw (0,0) -- (1,1) (0,1) -- (1,0);}$}%
\newcommand*{\router}[1]{
\begin{tikzpicture}
  \coordinate (ll) at (-3,0.5);
  \coordinate (lr) at (3,0.5);
  \coordinate (ul) at (-3,2);
  \coordinate (ur) at (3,2);
  \shade [shading angle=90, left color=switch, right color=white] (ll)
    arc (-180:-60:3cm and .75cm) -- +(0,1.5) arc (-60:-180:3cm and .75cm)
    -- cycle;
  \shade [shading angle=270, right color=switch, left color=white!50] (lr)
    arc (0:-60:3cm and .75cm) -- +(0,1.5) arc (-60:0:3cm and .75cm) -- cycle;
  \draw [thick] (ll) arc (-180:0:3cm and .75cm)
    -- (ur) arc (0:-180:3cm and .75cm) -- cycle;
  \draw [thick, shade, upper left=switch, lower left=switch,
    upper right=switch, lower right=white] (ul)
    arc (-180:180:3cm and .75cm);
  \node at (0,0.5){\color{blue!60!black}\Huge #1};
  \begin{scope}[yshift=2cm, yscale=0.28, transform shape]
    \node[route, rotate=45, xshift=\shift] {\strut};
    \node[route, rotate=-45, xshift=-\shift] {\strut};
    \node[route, rotate=-135, xshift=\shift] {\strut};
    \node[route, rotate=135, xshift=-\shift] {\strut};
  \end{scope}
\end{tikzpicture}}

\makeatletter
\pgfdeclareradialshading[tikz@ball]{cloud}{\pgfpoint{-0.275cm}{0.4cm}}{%
  color(0cm)=(tikz@ball!75!white);
  color(0.1cm)=(tikz@ball!85!white);
  color(0.2cm)=(tikz@ball!95!white);
  color(0.7cm)=(tikz@ball!89!black);
  color(1cm)=(tikz@ball!75!black)
}
\tikzoption{cloud color}{\pgfutil@colorlet{tikz@ball}{#1}%
  \def\tikz@shading{cloud}\tikz@addmode{\tikz@mode@shadetrue}}
\makeatother

\tikzset{my cloud/.style={
     cloud, draw, aspect=2,
     cloud color={gray!5!white}
  }
}

\renewcommand\thetable{\arabic{table}}
\usepackage{etoolbox}
\makeatletter
\patchcmd{\@makecaption}
  {\scshape}
  {}
  {}
  {}
\makeatother

\IEEEoverridecommandlockouts\IEEEpubid{\makebox[\columnwidth]{978-1-6654-0601-7/22/\$31.00 ~\copyright~2022 IEEE \hfill} \hspace{\columnsep}\makebox[\columnwidth]{ }}

\begin{document}
\bstctlcite{MyBSTcontrol}
\title{A System for Interactive Examination \\of Learned Security Policies}

\author{\IEEEauthorblockN{Kim Hammar \IEEEauthorrefmark{2}\IEEEauthorrefmark{3} and Rolf Stadler\IEEEauthorrefmark{2}\IEEEauthorrefmark{3}}

 \IEEEauthorblockA{\IEEEauthorrefmark{2}
Division of Network and Systems Engineering, KTH Royal Institute of Technology, Sweden
 }\\
 \IEEEauthorblockA{\IEEEauthorrefmark{3} KTH Center for Cyber Defense and Information Security, Sweden \\
Email: \{kimham, stadler\}@kth.se%
\\
\today
}
}

\maketitle
\begin{abstract}
We present a system for interactive examination of learned security policies. It allows a user to traverse episodes of Markov decision processes in a controlled manner and to track the actions triggered by security policies. Similar to a software debugger, a user can continue or or halt an episode at any time step and inspect parameters and probability distributions of interest. The system enables insight into the structure of a given policy and in the behavior of a policy in edge cases.
We demonstrate the system with a network intrusion use case. We examine the evolution of an IT infrastructure's state and the actions prescribed by security policies while an attack occurs. The policies for the demonstration have been obtained through a reinforcement learning approach that includes a simulation system where policies are incrementally learned and an emulation system that produces statistics that drive the simulation runs.

\end{abstract}

\begin{IEEEkeywords}
Network security, automation, reinforcement learning, Markov decision processes, MDP, POMDP.
\end{IEEEkeywords}

\IEEEpeerreviewmaketitle
\section{Introduction}
An organization's security strategy has traditionally been defined, implemented, and updated by domain experts \cite{int_prevention}. Although this approach can provide basic security for an organization's communication and computing infrastructure, a growing concern is that infrastructure update cycles become shorter and attacks increase in sophistication. Consequently, the security requirements become increasingly difficult to meet. To address this challenge, significant efforts have started to automate security frameworks and the process of obtaining effective security policies. Examples of this research include: computation of security policies using dynamic programming and control theory; computation of exploits and corresponding defenses through evolutionary methods; computation of security policies through game-theoretic methods; and use of machine learning techniques to estimate model parameters and policies \cite{ai_survey,hammar_stadler_cnsm_20, hammar_stadler_cnsm_21,hammar_stadler_tnsm21}.

A promising direction of recent research is automatically learning security policies through reinforcement learning methods (see survey \cite{ai_survey}). This paper builds on our earlier work in this area \cite{hammar_stadler_cnsm_20, hammar_stadler_cnsm_21,hammar_stadler_tnsm21}. In this line of research, the problem of finding a security policy is generally modeled as a Markov decision problem and policies are learned through simulation. Following this approach, the learned policies are evaluated by metrics that are aggregated over a large number of simulation runs. This kind of of evaluation provides a useful, overall, assessment of the policies' performance, but it is incomplete. For instance, it does not explain the structural differences between effective and non-effective policies. Also, it does not give insight into the behavior of policies in edge-cases, which is vital to understand before deploying security policies in practice.


In this demo paper, we describe a system that complements quantitative evaluations of learned security policies by allowing a user to interactively inspect the behavior of policies in different states and against different attackers. We demonstrate that the system enables insight into the structure of a given policy and in the behavior of a policy in edge cases \footnote{Video: \url{https://www.youtube.com/watch?v=18P7MjPKNDg&t=1s}}. We also open-source the code \cite{github_dashboard_hammar_stadler}, explain the implementation, and describe our reinforcement learning approach for finding security policies.

\begin{figure}
  \centering
  \scalebox{0.93}{
    \input{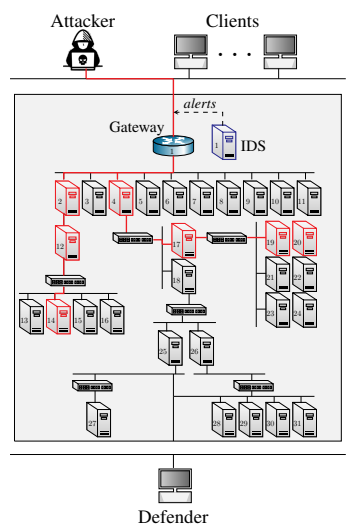}
    }
    \caption{The IT infrastructure and the actors in the intrusion prevention use case: an attacker, a client population, and a defender.}
    \label{fig:system2}
  \end{figure}
\section{The Intrusion Prevention Use Case}\label{sec:example_use_case}
We demonstrate the system with an \textit{intrusion prevention} use case (see Fig. \ref{fig:system2}). Here, we use the term "intrusion prevention'' as suggested in the literature, e.g. in \cite{int_prevention}. It means that a defender prevents an attacker from reaching its goal, rather than preventing it from accessing any part of the infrastructure.
\begin{figure}
  \centering
  \scalebox{0.95}{
    \input{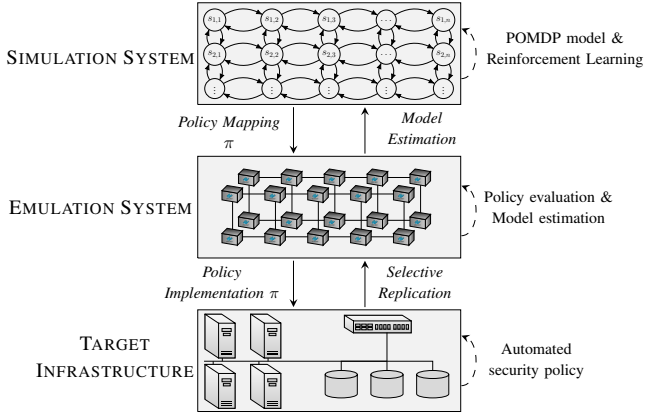}
 }
    \caption{Our framework for learning and evaluating security policies.}
    \label{fig:method}
\end{figure}

The use case involves the IT infrastructure of an organization (see Fig. \ref{fig:system2}). The operator of this infrastructure, which we call the defender, takes measures to protect it against an attacker while, at the same time, providing a service to a client population. The infrastructure includes a set of servers that run the service and an intrusion detection system (IDS) that logs events in real-time. Clients access the service through a public gateway, which also is open to the attacker.

We assume that the attacker intrudes into the infrastructure through the gateway, performs reconnaissance, and exploits found vulnerabilities, while the defender continuously monitors the infrastructure through accessing and analyzing IDS statistics and login attempts at the servers.

In our demonstration, we assume that the defender can take a fixed number of actions, each of which has a cost. An example of such an action is to update the firewall of the gateway. When deciding on an action, the defender balances two objectives: (\textit{i}) maintaining service to its clients; and (\textit{ii}), keeping a possible attacker out of the infrastructure.

For simulation of the use case and formal analysis, we model it with a Partially Observed Markov Decision Process (POMDP) \cite{hammar_stadler_tnsm21,hammar_stadler_cnsm_21}. The POMDP evolves in discrete time-steps: $t=1,\hdots,T$. The states $s_t \in \mathcal{S}$ and observations $o_t \in \mathcal{O}$ of the POMDP represent the infrastructure's states and the defender's observations, respectively. The defender policy $\pi_{\theta}(a_t|h_t)$ determines the actions $a_t \in \mathcal{A}$ that the defender takes after observing the history $h_t=(\rho_1,a_1,o_1,\hdots,a_{t-1},o_t)$, where $\rho_1$ is the initial state distribution.
\section{Automated Learning of Security Policies}\label{sec:obtaining_policies}
We integrate the system for interactive examination of learned security policies with a reinforcement learning framework for learning security policies that we developed in our previous work \cite{hammar_stadler_cnsm_20,hammar_stadler_tnsm21,hammar_stadler_cnsm_21}. Our framework includes two systems (see Fig. \ref{fig:method}). First, we use an \textit{emulation system} where key functional components of the target infrastructure are replicated. This system closely approximates the functionality of the target infrastructure and is used to run attack scenarios and defender responses. These runs produce system metrics and logs that we use to estimate distributions of infrastructure metrics. We then use the estimated distributions to instantiate a POMDP. Second, we use a \textit{simulation system} where POMDP episodes are simulated and policies are incrementally learned. After learning policies, they are extracted from the simulation system and evaluated in the emulation system. Using this approach, the emulation system provides the statistics needed to simulate the use case and to evaluate the learned policies, whereas the simulation system automatically learns policies through reinforcement learning in an efficient manner.
 \section{The System for Interactive Policy Examination}
\begin{figure}
  \centering
  \scalebox{1.1}{
    \input{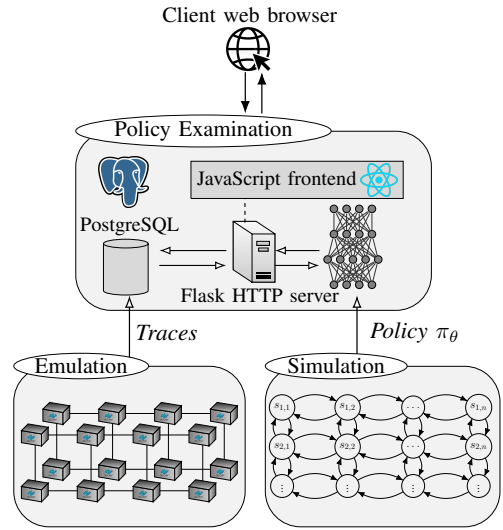}
 }
    \caption{Architecture of the system for inspecting and examining learned security policies.}
    \label{fig:impl}
  \end{figure}

The architecture of the system for policy examination is shown in Fig. \ref{fig:impl}. The system provides an interface to inspect learned policies either from emulation traces or from simulating a POMDP. Users access the system through a web browser where they can interactively step through a POMDP episode. At each step, the users can inspect relevant parameters and probability distributions.

\begin{figure*}
  \centering
    \scalebox{0.40}{
      \includegraphics{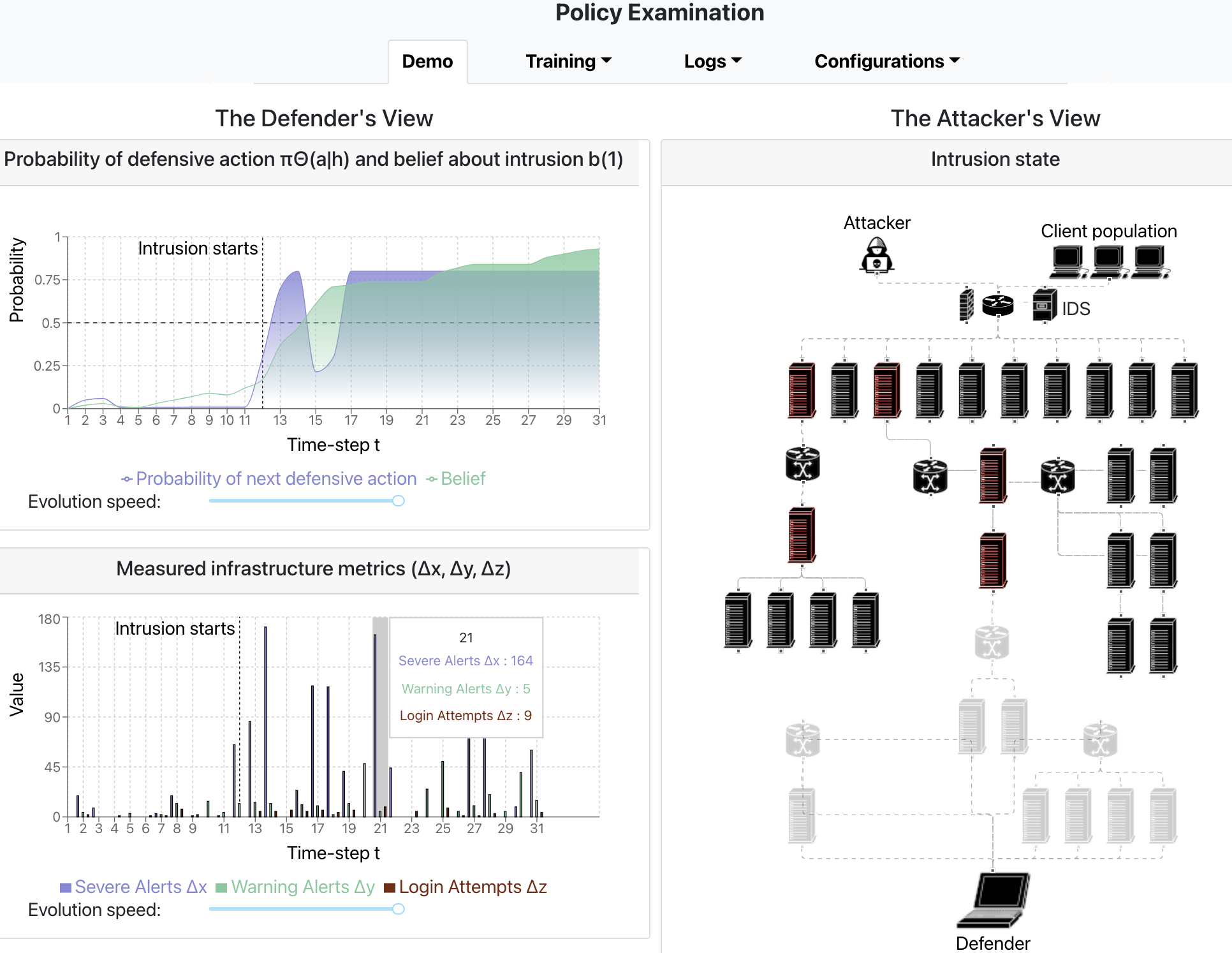}
    }
    \caption{The demonstration view of the system which allows the user to interactively step through a POMDP episode.}
    \label{fig:demo_1}
  \end{figure*}
\section{Implementation}
The web interface is implemented in JavaScript using the React framework (see Fig. \ref{fig:impl}). It makes requests to an HTTP server written in Python based on the Flask framework, which provides a REST API that can be queried to fetch data and to make policy predictions. The data on the server is collected from the emulation system and is stored in a PostgreSQL database. The policy is extracted from the simulation system and is stored in the main memory of the server.

The emulation system executes on a cluster of machines that runs a virtualization layer provided by Docker \cite{docker} containers and virtual connections. It implements network isolation and traffic shaping on the containers using network namespaces and the NetEm module in the Linux kernel. Resource constraints of the containers, e.g. CPU and memory constraints, are enforced using cgroups. The software running in the containers replicates important components of the target infrastructure, such as, web servers, databases, and an IDS.

The defender policy  $\pi_{\theta}(a|h)$ is represented by a neural network that takes infrastructure metrics $h$ as input and outputs a probability distribution over actions $a \in \mathcal{A}$.

The simulation system is written in Python, and the policy training is performed using the Proximal Policy Optimization (PPO) algorithm, which we have implemented using the PyTorch auto-differentiation library.
\section{Demonstration}
We demonstrate the system by interactively stepping through a POMDP episode of the intrusion prevention use case. The POMDP is traversed based on traces and a defender policy, which were obtained through the reinforcement learning method described in Section \ref{sec:obtaining_policies}.

Figure \ref{fig:demo_1} pictures the main user interface of the system. The left part of the interface shows the defender's view and the right part shows the attacker's view. The plot on the upper left shows the defender's belief about the infrastructure's state and the probability that the defender takes an action at each time-step. The plot on the lower left shows the distribution of infrastructure metrics that the defender observes. The graphic on the right shows the attacker's view, depicted as an overlay on the IT-infrastructure's topology.

The demonstration provides insight into the structure of the defender policy and its behavior in edge-cases. We observe correlations among the attacker's actions, the infrastructure metrics, and the actions that the defender policy prescribes. We also examine which actions of the attacker are difficult for the defender to detect or trigger actions by the defender.

For example, we can observe that if the attacker performs reconnaissance by executing a port scan, there is a spike in infrastructure metrics, which causes a defensive action with a high probability. If the attacker uses other means of reconnaissance, e.g. a ping scan, we can observe that the defender will not detect it since no change in infrastructure metrics is generated. Finally, we can discover an edge-case where the defender policy mistakes regular client traffic for an attacker and wrongly takes defensive actions. This happens, for instance, if the attacker is passive for a long time.


\ifCLASSOPTIONcaptionsoff
  \newpage
\fi

\bibliographystyle{IEEEtran}
\bibliography{references,url}

\end{document}